\documentclass[sigconf,
                nonacm,
                review=False,
                anonymous=false
        ]{acmart}
\usepackage{amsmath}
\usepackage{graphicx}
\usepackage{amsfonts}
\usepackage{amsthm}
\usepackage{booktabs}
\usepackage[ruled,vlined,linesnumbered]{algorithm2e}
\usepackage{dsfont}
\usepackage{pifont}
\usepackage{algorithmic}
\AtBeginDocument{%
  }

\setcopyright{acmlicensed}
\copyrightyear{2024}
\acmYear{2024}
\acmDOI{XXXXXXX.XXXXXXX}

\acmConference[ICAIF '24]{International Conference on AI in Finance}{November 14--16,
  2024}{Brooklyn, NY}
\acmISBN{978-1-4503-XXXX-X/18/06}

\begin{document}


\title[Evaluating Financial Relational Graphs]{Evaluating Financial Relational Graphs: Interpretation Before Prediction}

\author{Yingjie Niu}
\orcid{0000-0001-9322-2726}
\affiliation{%
  \institution{School of Computer Science, University College Dublin}
  \city{Dublin}
  \country{Ireland}
}
\email{yingjie.niu@ucdconnect.ie}

\author{Lanxin Lu}
\orcid{0000-0003-3271-2750}
\affiliation{%
  \institution{Michael Smurfit Business School, University College Dublin}
  \city{Dublin}
  \country{Ireland}
}
\email{lanxin.lu@ucdconnect.ie}

\author{Rian Dolphin}
\orcid{0000-0002-5607-9948}
\affiliation{%
  \institution{School of Computer Science, University College Dublin}
  \city{Dublin}
  \country{Ireland}
}
\email{rian.dolphin@ucdconnect.ie}

\author{Valerio Poti}
\orcid{0000-0003-1156-5616}
\affiliation{%
  \institution{Michael Smurfit Business School, University College Dublin}
  \city{Dublin}
  \country{Ireland}
}
\email{valerio.poti@ucd.ie}

\author{Ruihai Dong}
\orcid{0000-0002-2509-1370}
\affiliation{%
  \institution{School of Computer Science, University College Dublin}
  \city{Dublin}
  \country{Ireland}
}
\email{ruihai.dong@ucd.ie}


\begin{abstract}
Accurate and robust stock trend forecasting has been a crucial and challenging task, as stock price changes are influenced by multiple factors. Graph neural network-based methods have recently achieved remarkable success in this domain by constructing stock relationship graphs that reflect internal factors and relationships between stocks. However, most of these methods rely on predefined factors to construct static stock relationship graphs due to the lack of suitable datasets, failing to capture the dynamic changes in stock relationships. Moreover, the evaluation of relationship graphs in these methods is often tied to the performance of neural network models on downstream tasks, leading to confusion and imprecision. To address these issues, we introduce the SPNews dataset, collected based on S\&P 500 Index stocks, to facilitate the construction of dynamic relationship graphs. Furthermore, we propose a novel set of financial relationship graph evaluation methods that are independent of downstream tasks. By using the relationship graph to explain historical financial phenomena, we assess its validity before constructing a graph neural network, ensuring the graph's effectiveness in capturing relevant financial relationships. Experimental results demonstrate that our evaluation methods can effectively differentiate between various financial relationship graphs, yielding more interpretable results compared to traditional approaches. We make our source code publicly available on GitHub to promote reproducibility and further research in this area \footnote{https://github.com/FreddieNIU/Financial-Graph-Evaluation}.

\end{abstract}

\ccsdesc[500]{Computing methodologies~Machine learning}
\ccsdesc[500]{Computing methodologies~Knowledge representation and reasoning}

\keywords{Representation Learning, Financial Markets, Graph Neural Network, Graph Evaluation, Artificial Intelligence, Machine Learning}


\maketitle

\section{Introduction}
Stock market prediction using machine learning techniques has garnered significant attention in recent years \cite{li2016tensor,ding2016knowledge,li2020multimodal}. 
Aside from being influenced by its own momentum, a stock's price is also affected by the momentum spillovers among related companies, indicating that the dynamics of related firms' stock prices can exert an impact \cite{ali2020shared}. 
The advent of graph neural networks (GNNs) has enabled researchers to model these momentum spillover effects by constructing corporate relationship graphs and training GNN models on them \cite{chen2018incorporating,li2021modeling}. However, most prior works rely on predefined, static relationships to construct these graphs, which may not capture the dynamic nature of inter-company relationships in fast-paced stock markets. Predefined relationships may become outdated and no longer applicable to the current market. Some recent studies have explored building dynamic relationship graphs using historical market signals~\cite{xiang2022temporal}. Still, due to data limitations, these graphs are often constructed solely from quantitative data without leveraging alternative data that could help define company relationships. News, as a high-frequently updated information source, can serve as a good resource for capturing the changing of corporate relationships. Thus, there is a need for a news dataset that can assist dynamic relationship graph construction. To solve this problem we introduce the SPNews dataset, consisting of a news dataset collected based on S\&P500 Index Stocks. This dataset will help future researchers explore more possibilities for building dynamic relationship graphs that incorporate alternative data.

Moreover, existing approaches for evaluating relationship graphs primarily rely on their performance on downstream tasks, which can be misleading. When concurrently appraising graphs and graph neural network models, a robust model might obscure deficiencies within the constructed graph. 
Furthermore, another issue arising from selecting graphs based on downstream task performance is the limited generalization ability of such graphs. Graphs selected in this manner may only prove effective for the specific task or even solely within the confines of the dataset used for selection. Thus, we claim that: selecting the graph solely based on its performance in downstream tasks amounts to conflating the evaluating the graph itself (an upstream task) and assessing its suitability for downstream tasks. Keeping these tasks separate not only enhances the interpretability of the graph but also improves its generalization ability, because a graph that yields satisfactory results in a limited dataset may not generalize well if the collinearities among nodes are unstable over time. 

In this work, we introduce a novel financial graph evaluation framework designed to \emph{decouple the evaluation of relationship graphs from downstream tasks}. The framework operates independently of neural network models or downstream tasks. We argue that, prior to the training of a graph neural network derived from a relationship graph, a comprehensive evaluation of the relationship graph itself is imperative. Our central tenet posits that interpretation should take precedence over prediction. 
Experiment results prove that our method can effectively evaluate the differences between financial relationship graphs, and the results of our evaluation method are more interpretable than traditional methods. We summarize our contributions as follows: 
\begin{itemize}
\item We release a news dataset collected based on SP500 Index Stocks which benefits the dynamic firm relationship graph construction.
\item We propose an interpretable graph evaluation framework that operates independently of neural network models or downstream tasks.
\item Through experiments on various relationship graphs, we prove that the proposed evaluation framework can effectively assess different relationship graphs.
\end{itemize}

\section{Related Work}
Statistical and machine learning techniques have been widely applied to stock trend forecasting, leveraging both time series data and alternative data sources. Traditional models, such as the Autoregressive Integrated Moving Average (ARIMA) model, have long been used to capture the temporal dependencies in stock returns~\cite{ariyo2014stock}, while more recently, deep learning models, such as Long Short-Term Memory (LSTM) networks \cite{hochreiter1997long}, have demonstrated strong performance in capturing complex patterns and non-linear relationships in financial time series data~\cite{sezer2020financial}. Many researchers explored the possibility of applying Natural Language Processing (NLP) techniques in the financial field resulting in notable success \cite{duan-etal-2018-learning,niu2023learning}. In addition to time series data, alternative data sources such as news articles \cite{ding2015deep}, social media posts \cite{xu2018stock,sawhney2020deep}, company filings, and audio data from earnings calls \cite{yang2020html} have been utilized to extract relevant features for stock price prediction. 

With recent advances in graph-based machine learning and representation learning, relational information within financial markets is being explored in more detail to move away from treating assets independently~\cite{dolphin2022stock}. Researchers start to model the momentum spillover effect through corporate relationship graphs and graph neural networks (GNN) \cite{scarselli2008graph,xu2022hgnn}. More advanced GNN-based techniques like Graph Attention Networks (GAT) \cite{velivckovic2017graph}, for example, have recently gained traction in stock returns forecasting, as they can effectively capture the complex relationships and dependencies among stocks~\cite{ding2016knowledge,li2021modeling,lei2024dr}. Among these studies, a common practice in graph construction is that they build a static corporate relation graph based on predefined relations.  Few attempts have been made to capture dynamic relationships based on the correlation of historical data \cite{xiang2022temporal}. 

Moreover, the evaluation of financial relationship graphs remains a challenging task, with most existing approaches relying on the performance of graph-based models on downstream tasks~\cite{li2021modeling,cheng2021modeling,xiang2022temporal}. There is a need for more comprehensive and standardized evaluation frameworks that can assess the quality of financial relationship graphs independently of their application in downstream tasks. We shrink this gap by proposing a novel interpretable framework for evaluating financial relationship graphs.

\begin{table}[]
\begin{tabular}{ll}
\hline
Symbol                                              & Definition                                       \\ \hline
$\mathcal{G} = \{\mathcal{G}_0,... \mathcal{G}_T\}$ & dynamic relationship graph set               \\
$\mathcal{G}_t = (\mathcal{V}, E_t) $                & relationship graph at timestamp $t$          \\
$T$                                                 & number of trading days of $\mathcal{G}$      \\
$\mathcal{V}_t^m$                                   & set of nodes with degree \textgreater{}=1 at $t$   \\
$\mathcal{V}^M = \mathcal{V}_0^m \bigcap ... \mathcal{V}_T^m$                                   & set of nodes with degree \textgreater{}=1   \\
$\mathcal{V}_t^n$                                   & set of nodes with degree \textless{}1 at $t$       \\
$\mathcal{V} = \{\mathcal{V}_t^m, \mathcal{V}_t^n\}$    & the set of all nodes                             \\
$m_t$                                                 & number of nodes in $\mathcal{V}_t^m$ \\
$n_t$                                                 & number of nodes in $\mathcal{V}_t^n$    \\
$M$                                                 & number of node pairs with edges \\
$E_t$                                               & set of edges in $\mathcal{G}_t$             \\ 
$\sigma(\cdot)$                                                 & the correlation calculation function \\
$\mu$                                                 & the edge index \\
$\nu$                                                 & the edge attribute (strength) \\ 
$\epsilon$                                                 & rolling window length \\ \hline
\end{tabular}
\caption{Mathmatical Symbols Summary}
\label{tab:symbol}
\end{table}

\section{Problem Definition and Graph Construction}
In this section, we conceptualize the financial relationship graphs and introduce the way we construct the dynamic relationship graph set $\mathcal{G}$ based on our SPNews dataset.

\begin{figure*}[]  
    \centering
    \includegraphics[scale = 0.115]{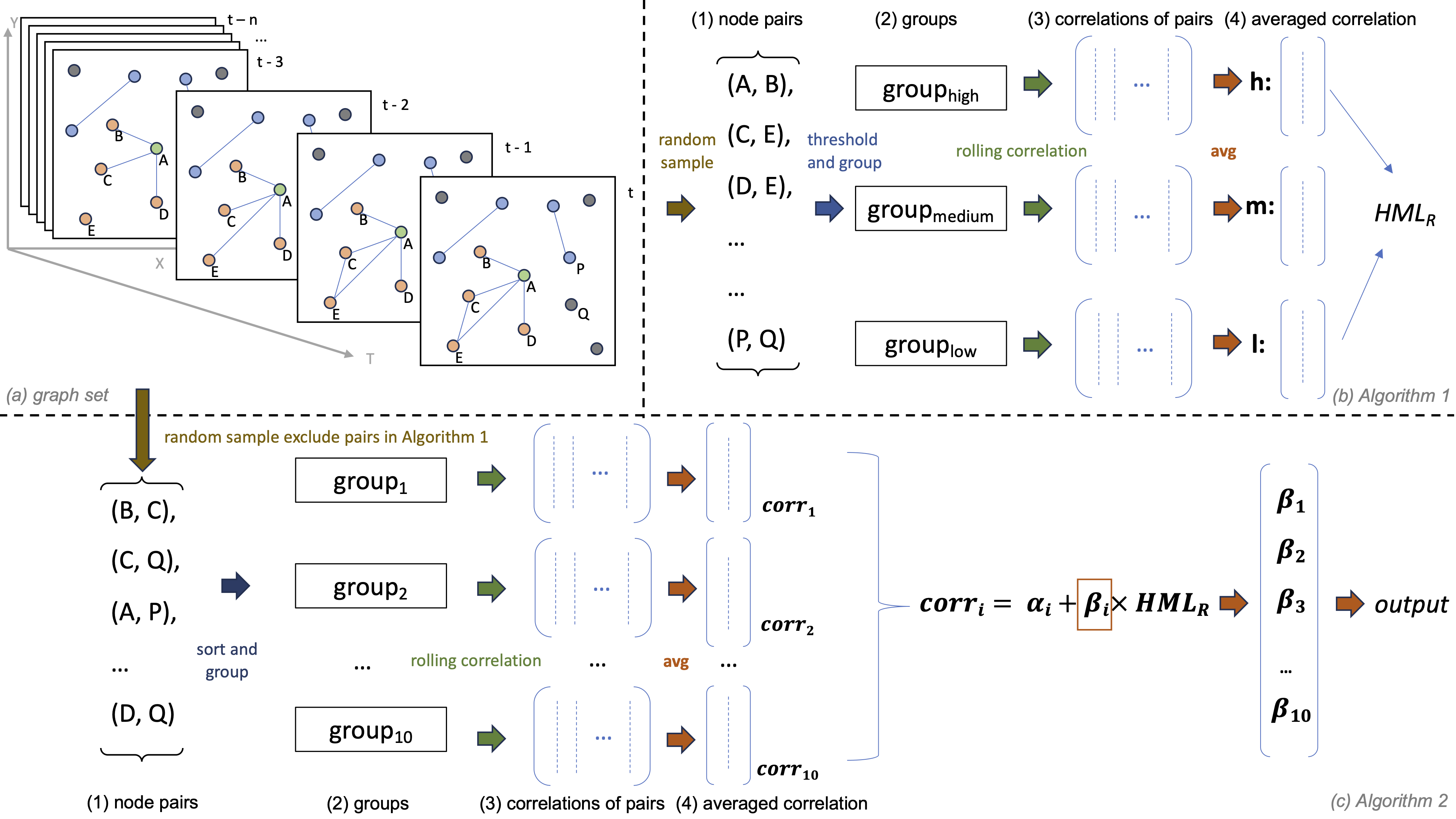}
    \caption{(a): Dynamic relationship graph set $\mathcal{G}$. (b): relationship factor ($HML_R$) construction. (c): $HML_R$ evaluation. (b) and (c) corresponds to Algorithm \ref{alg:factor-construct} \& \ref{alg:factor-test} respectively. A, B, C, D, P, and Q are examples of nodes(companies). In each matrix of (b)(3) and (c)(3), each column is the rolling correlation coefficient of a node pair. The 3 matrices in (b)(3) have different shapes, that are $[T-\epsilon, n_{high}], [T-\epsilon, n_{medium}], [T-\epsilon, n_{low}]$ respectively, where $n_{high}, n_{medium}, n_{low}$ represent the number of node pairs in $group_{high}, group_{medium}$, and $group_{low}$. $T$ and $\epsilon$ follow the definition in Table \ref{tab:symbol}. The 10 matrices in (c)(3) have the same shape $[T-\epsilon, 100]$.}
    \label{fig:algorithms}
\end{figure*}

\textbf{Problem Definition} In contrast to conventional graph-based methodologies, which manually create static graphs through pre-defined relationships, we conceptualize the company relation graph set as an assemblage of temporal evolutional graphs. Within these graphs, each node signifies a firm, and the edges encapsulate their relations. Figure \ref{fig:algorithms} (a) illustrates the temporal evolution of a company relationship graph on a three-dimensional coordinate axis. Each X-Y plane corresponds to a specific relationship graph $\mathcal{G}_t$ at time $t$. The T-axis signifies time, capturing the progressive changes in the relationship graph over temporal intervals. And we name all the relationship graphs within a certain period $T$ as a relationship graph set $\mathcal{G}$.

\textbf{Graph Construction} Table \ref{tab:symbol} summarizes the symbols introduced in this paper. The graph set $\mathcal{G}$ consists of a series of relationship graphs at different timestamps. A relationship graph $\mathcal{G}_t = (\mathcal{V}, E_t)$ where $\mathcal{V}$ is the set of nodes which is constant through the whole period, and $E_t$ is the set of edges in $\mathcal{G}_t$. Let A and B represent a pair of nodes, the edge between A and B at time $t$ is represented as $E_t(A,B) = (\mu_t^{(A,B)}, \nu_t^{(A,B)})$ where $\mu_t^{(A,B)}$ is a boolean value indicates whether there exists an edge between A and B at time $t$. $\nu_t^{(A,B)}$ is a float number attached to an edge to record the strength of the connection. To construct an edge $E_t(A,B)$ based on the SPNews dataset, we look for news that mentioned both A and B among all the news on day t. If such news exists, we count the number of these news as $k$ and compare $k$ with a pre-defined threshold $\tau$. If $k > \tau$,  we assign the edge $E_t(A,B) = (1, Norm_{\mathcal{G}_t}(k))$, otherwise, $E_t(A,B) = (0, 0)$. $Norm_{\mathcal{G}_t}(k)$ indicates that we normalize the edge attribute within each graph $\mathcal{G}_t$. 

\section{Graph Evaluation Methods}

In this section, we introduce the proposed graphs evaluation methods, named \textit{Financial Relationship-graph Interpretation (FRI)} framework which contains 4 indicators. We utilize the graph to interpret inter-company financial phenomena, substantiating the efficacy of the constructed relationship graph. 
To comprehensively evaluate the entire graph set $\mathcal{G}$, we conduct the evaluation on two dimensions:
\begin{itemize}
\item \textbf{Horizontal:} Within each graph $\mathcal{G}_t$, analyse and compare the difference between connected nodes $\mathcal{V}_t^m$ and isolated nodes $\mathcal{V}_t^n$ to interprate the relationships built at time $t$. 
\item \textbf{Vertical:} Along the T axis, for each pair of firms $A$ and $B$, analyse the change of their relationship (edges) along time evolution $[E_0(A,B), E_1(A,B), E_2(A,B), ..., E_T(A,B)]$. 
\end{itemize}

\subsection{Return Correlation Stability}
The correlation coefficient derived from stock historical return data typically serves as a measure of the relationship between the two firms. Denoting the correlation coefficient between firm A and firm B during the period $[t, t+\epsilon]$ as $\sigma_t^{t+\epsilon}(A, B)$, the alteration in correlation before and after $E_t(A, B)$ is established can be expressed as 
\begin{align}
    \delta_t(A,B) = \sigma_t^{t+\epsilon}(A,B) - \sigma_{t-\epsilon}^t(A,B) 
\end{align}%
where $\epsilon$ is the window length ($21$ trading days in our experiment) used to calculate the correlation coefficient. To evaluate the edges (relationships) within a relationship graph $\mathcal{G}_t$, we have the following assumptions: 1. the correlation between two companies without any relationship always fluctuates randomly. 2. The correlation between two companies with a relationship is usually affected by changes in their relationship. Therefore, by comparing the correlation changes before and after an edge is established, we can evaluate whether this edge effectively captures the true relationship and its changes.
We anticipate that changes in correlation between companies with established edges will be significantly larger than that between companies lacking such connections. Consequently, we propose the following null hypothesis.
\begin{itemize}
\item $\mathbf{H_0}$ The change in the correlation of connected nodes is lower than that of the non-connected nodes. 
\end{itemize}
$\mathbf{H_0}$ can be formulated mathematically as 
\begin{align}
    & |\delta_t(A_1,B_1)| <= |\delta_t(A_2,B_2)|
\end{align}%
where $A_1, B_1 \in \mathcal{V}_t^m$ and $A_2, B_2 \in \mathcal{V}_t^n$ with notation following the definitions in Table \ref{tab:symbol}.

By comparing the difference in correlation stability between connected nodes and unconnected nodes in $\mathcal{G}_t$, it can be demonstrated whether the relationship graph effectively captures the company pairs in which true relationships exist. We first test the null hypothesis on each $\mathcal{G}_t$ to calculate the \textit{Correlation Stability (CS)} on $\mathcal{G}_t$.
\begin{align}
CS_{\mathcal{G}_t}= 
\begin{cases}
    1,  & \text{if $\mathbf{H_0}$ is rejected} \\
    0,  & \text{otherwise}
\end{cases}
\end{align}%
Then we propose \textit{Correlation Stability Score (CSS)} to formulate the return correlation stability as a quantitative indicator of the entire graph set $\mathcal{G}$.
\begin{align}
    CSS = \frac{1}{T} \sum_{\mathcal{G}_t\in\mathcal{G}} CS_{\mathcal{G}_t} \nonumber\\
\end{align}%
where $T$ represents the number of trading days in $\mathcal{G}$, which is also the number of graphs in the graph set. As such, $CSS$ can be interpreted as the proportion of graph $\mathcal{G}_t$ where the correlation change of connected nodes is significantly greater than that of unconnected nodes.


\subsection{Event Detection}
In order to comprehensively evaluate a dynamic company relationship graph, it is not only necessary to conduct horizontal comparisons among companies within $\mathcal{G}_t$, but also to analyze how the relationship between two companies evolves over time, so as to evaluate whether the dynamic graph adequately captures the changes of the relationship between the two companies during this period. 

Through observation, it is found that two firms usually co-occur in many news articles during some periods, but not at all at other times. We define the periods that have continued co-occurrence of two firms as an \textit{event period}, that is $[day_t, ..., day_{t+T_e}]$ where $t$ is the starting date of the event period, and $T_e$ is the number of trading days in this period. In this notation, the period of the entire graph set $\mathcal{G}$ is $[day_0, ..., day_T]$. If a significant change in the correlation of the two firms is observed during the event period, it means some breaking events have happened which affects the correlation of the two firms. For example, if the correlation of firm A and firm B decreased from 0.7 to 0.3 during an event period, we can infer that some breaking events happened, which led to a drop in the correlation strength. The news articles associated with the edges between firms A and B built during that event period can explain the drop in correlation. In order to quantify how well a graph set $\mathcal{G}$ captures events, we propose the \textit{Average Event Capture Rate (AECR)}. Let firm A and firm B represent a pair of firms within the graph set $\mathcal{G}$, the maximum correlation difference of firm A and B over whole period is
\begin{align}
    \Delta_{T}(A,B) &= \max([\sigma_0^\epsilon(A,B),..., \sigma_{T-\epsilon}^T(A,B)]) - \nonumber\\
    & \min([\sigma_0^\epsilon(A,B),..., \sigma_{T-\epsilon}^T(A,B)])
\end{align}%
The maximum correlation difference over one \textit{event period} is
\begin{align}
    \Delta_{T_e}(A,B) &= \max([\sigma_t^{t+\epsilon}(A,B),..., \sigma_{t+T_e-\epsilon}^{t+T_e}(A,B)]) - \nonumber\\
    & \min([\sigma_t^{t+\epsilon}(A,B),..., \sigma_{t+T_e-\epsilon}^{t+T_e}(A,B)])
\end{align}%
where $t$ is the starting date of the event period. The event-capturing indicator $EC_{(A,B), T_e}$ of the event period is
\begin{align}
EC_{(A,B), T_e}= 
\begin{cases}
    1,  & \text{if } \frac{\Delta_{T_e}(A,B)}{\Delta_{T}(A,B)} > \mathrm{std}([\sigma_0^\epsilon(A,B),..., \\
    & \sigma_{T-\epsilon}^T(A,B)])\\
    0,  & \text{otherwise}
\end{cases}
\end{align}%
By averaging the $EC_{(A,B), T_e}$ over all of the event periods, the \textit{Event Capture Rate (ECR)} of the firm pair A and B is then calculated as
\begin{align}
    ECR_{(A,B)} = \frac{1}{\rho} \sum_{T_e}EC_{(A,B), T_e} 
\end{align}%
where $\rho$ represents the number of event periods between firm A and firm B. Then, the \textit{ECR} of each pair of firms is calculated, and the averaged value is used as a quality indicator of the graph set $\mathcal{G}$, which is named as \textit{Averaged Event Capture Rate (AECR)}.
\begin{align}
    AECR = \frac{1}{M}\sum_{A\in\mathcal{V}^M} \sum_{\substack{B\in\mathcal{V}^M\\A\neq B}} ECR_{(A,B)}
\end{align}%
where $\mathcal{V}^M$ and $M$ with notation following the definitions in Table \ref{tab:symbol}.

\subsection{Edge Factor Model: Explain Return Correlation}
The Fama-French Three Factor model is a formula to describe the rate of return on a stock investment \cite{fama1993common}. 
This model evaluates the anticipated rate of return on investment by considering three factors: overall market risk, the relative outperformance of small-cap over large-cap companies, and the extent to which high-value companies outperform low-value ones. Drawing inspiration from the Fama-French Three Factor model, we introduce a relationship factor, i.e. $HML_R$, to elucidate the return correlation or co-movement between two firms. Our assumption is that \textit{during period $[day_0, ..., day_T]$, the higher the density of edges established between two companies, the higher the correlation between the two companies}. The relationship factor construction process is demonstrated in Algorithm \ref{alg:factor-construct} and Figure \ref{fig:algorithms} (b). In the Algorithms, we use lower case letter(s) to represent scalar, bold lower case letter(s) to represent vector, and bold upper case letter(s) to represent matrix. 

To evaluate the effectiveness of the $HML_R$ factor, we conduct the test on a group of node pairs different from the 1200 pairs used in factor construction. The factor testing process is demonstrated in Algorithm \ref{alg:factor-test} and Figure \ref{fig:algorithms} (c) which returns a series of coefficient $\beta$. In the testing phase, node pairs were grouped based on the number of edges, and a regression analysis using the $HML_R$ factor was conducted on the return correlation within each group. If a noticeable upward trend is observed in the $\beta$ values, it serves as evidence supporting our assumption. Thus, we propose the averaged $\beta$ difference as the quantitative indicator of the explanatory capacity of $\mathcal{G}$ to the return correlation. The higher $\Delta_{\beta}$ indicates the better explanatory capacity.
\begin{align}
    \Delta_{\beta} = \frac{1}{h-1}\sum_{i=1}^{h-1} \left(\beta_{i+1} - \beta_i\right)
\end{align}%
where $h$ represents the number of groups in Algorithm \ref{alg:factor-test}.

From the financial graph evaluation point of view, a relationship factor constructed based on edges embedded in the graphs can effectively explain the differences in return correlation between companies, which substantiates the effectiveness of the edges contained in the graphs.

\begin{algorithm}[tb]
\caption{Relationship Factor Construct}
\label{alg:factor-construct}
\textbf{Input}: Dynamic Relationship Graph Set $\mathcal{G}$\\
\textbf{Parameter}: high boundary $\phi_h = 0.7$, low boundary $\phi_l = 0.3$\\
\textbf{Output}: Relationship Factor $HML_R$ \\
\begin{algorithmic}[1] 
\STATE Randomly sample 1200 node pairs from $G$, named $pairs$
\STATE Find the pair with the maximum number of edges and save the maximum edge number to the variable $max$
\STATE $h = \phi_h \times max$, $l = \phi_l \times max$
\STATE Loop through each pair and assign the pair to $group_{high}$, $group_{medium}$, or $group_{low}$ by comparing $NumOfEdges(pair)$ and $h, l$.
\STATE Within each group, calculate the rolling correlation coefficient of each pair of nodes as a time series. 
\STATE Average through node pairs within each group, gives us three time series: $\mathbf{h}, \mathbf{m}, \mathbf{l}$ correspond to the high correlated group, medium correlated group and low correlated group, respectively.  
\STATE \textbf{return} $HML_R = \mathbf{h} - \mathbf{l} $
\end{algorithmic}
\end{algorithm}

\begin{table*}[t]
\centering
\begin{tabular}{lllll}
\hline
Dataset                   & Period            & Stock Index      & Timestamp         & Mentioned Stock Label \\
\hline
Reuters \& Bloomberg 2014 & 10/2006 - 11/2013 & /                & Yes               & Not labelled          \\
Reuters 2018              & 10/2006 - 12/2015 & /                & No                & Not labelled          \\
Reuters 2021              & 01/2013 - 09/2018 & TPX500 \& TPX100 & Yes               & Partially labelled   \\
SPNews                    & 09/2022 - 10/2023 & SP500            & Yes               & Fully labelled        \\
\hline
\end{tabular}
\caption{Financial News Dataset Comparison. "Mentioned Stock Label" means whether there are labels of the news mentioned companies in each record.}
\label{tab:dataset}
\end{table*}

\subsection{Edge Factor Model: Explain Volatility Correlation}

The $HML_R$ factor evaluates the ability of the company relationship graph set $\mathcal{G}$ to explain the relationship between the rate of return among its nodes, which is very helpful for tasks such as stock trend prediction. However, relationship graphs can be widely applied to a variety of downstream tasks, and it is one-sided to only focus on the relationship between returns. The Dynamic Conditional Correlation Generalized Autoregressive Conditional Heteroscedasticity (DCC-GARCH) model \cite{engle2001theoretical} was introduced as an extension of the CCC-GARCH model \cite{bollerslev1990modelling} which focuses on modelling the volatility of individual financial time series. Therefore, we include the DCC-GARCH model in our assessment instruments that focus on the volatility correlation between firms. The implementation of DCC-GARCH model in this paper follows \cite{brownlees2017srisk}.

The evaluation of the dynamic relationship graph utilizing the DCC-GARCH model commences with the execution of steps 1 through 17 as delineated in Algorithm \ref{alg:factor-construct}, culminating in the categorization of company pairs into three distinct groups. Specifically, $group_{high}$ comprises node pairs exhibiting a notably strong correlation within the relational graph $\mathcal{G}$, akin to $group_{medium}$ and $group_{low}$ denoting varying degrees of relational strength. Subsequently, the DCC-GARCH model is applied to the returns of each node pair, yielding coefficients denoted as $\alpha$ and $\beta$. Averaging these coefficients within each group yields six group-level outcomes: $\alpha_{high}, \beta_{high}, \alpha_{medium}, \beta_{medium}, \alpha_{low}, \beta_{low}$. Within the context of DCC-GARCH results, the condition $\alpha + \beta < 1$ denotes model stability, signifying the efficacy of the dynamic correlation relationship. Here, $\alpha$ represents the degree of influence of residuals on the correlation coefficients, which in economic terms means the degree of influence of new information on the correlation of market volatility. $\beta$ represents the degree of influence of past market volatility on current market volatility, that is, the persistence degree of market volatility correlation. Thus, we proposed the $\Delta_{DCC}$ as an indicator: 
\begin{align}
    \Delta_{DCC} = \alpha_{high} - \alpha_{low} + \beta_{low} - \beta_{high}
\end{align}%
where $\alpha_{high}$ and $ \alpha_{low}$ indicate the averaged $\alpha$ within the $group_{high}$ and $group_{low}$. $\beta_{high}$ and $\beta_{low}$ represent the averaged $\beta$ value within $group_{high}$ and $group_{low}$. A positive value of $\alpha_{high}-\alpha_{low}$ signifies that the volatility correlation among company pairs in $group_{high}$ is more responsive to new information compared to that in $group_{low}$. Conversely, a positive value of $\beta_{low}-\beta_{high}$ suggests that the volatility correlation among company pairs in $group_{low}$ is more influenced by past volatility correlations than those in $group_{high}$. Thus, a higher $\Delta_{DCC}$ value indicates a greater discriminatory capacity of the graph set $\mathcal{G}$ in identifying firms with higher volatility correlation.

\begin{algorithm}[tb]
\caption{Relationship Factor Test}
\label{alg:factor-test}
\textbf{Input}: Dynamic Relationship Graph Set $\mathcal{G}$\\
\textbf{Parameter}: $HML_R$\\
\textbf{Output}: A series of $\boldsymbol{\beta}$ \\
\begin{algorithmic}[1] 
\STATE Randomly sample 1000 node pairs from $\mathcal{G}$ excluding pairs used in factor construction, named $pairs$
\STATE $pairs_{sorted}$ = sort $pairs$ by $NumOfEdges(p)$ in ascending order.
\STATE Split $pairs_{sorted}$ into 10 groups in order, each group has an equal number of pairs.
\STATE Within each group, calculate the rolling correlation coefficient
of each pair of nodes as a time series.
\STATE Average through node pairs within each group, gives us 10 time series: $\mathbf{CORR} = [\mathbf{corr_{1}}, \mathbf{corr_{2}}, ..., \mathbf{corr_{10}}]$.
\STATE $\boldsymbol{\beta} = []$
\FOR{$\mathbf{corr_{i}}$ in $\mathbf{CORR}$}
\STATE $\mathbf{corr_{i}} = \alpha_i + \beta_i \times HML_R$
\STATE $\boldsymbol{\beta}$ append $\beta_i$
\ENDFOR
\STATE \textbf{return} $\boldsymbol{\beta}$
\end{algorithmic}
\end{algorithm}

\section{Data Collection}
In this section, we introduce our SPNews dataset and discuss the advantages of our dataset compared to the existing financial news datasets. 
\subsection{Existing Financial News Dataset}
Business news is sourced from various outlets. This work leverages open-source datasets mentioned in prior research for comparative analysis.

\textbf{Reuters \& Bloomberg 2014}: A large-scale financial news dataset from Reuters and Bloomberg released in 2014 \cite{ding2014using}. 

\textbf{Reuters 2018}: A publicly available financial news dataset collected from Reuters from October 2006 to December 2015 \cite{duan-etal-2018-learning}. 

\textbf{Reuters 2021}: A financial news and market dataset collected from Reuters from January 2013 to September 2018 for the Tokyo Stock Exchange (TSE) \cite{li2021modeling}. 

\subsection{SPNews Dataset Description}
In this work, we choose stocks within the SP500 index, collect publicly available financial news articles from Yahoo Finance\footnote{https://finance.yahoo.com/} during September 2022 and October 2023, and publish our dataset named SPNews. We omit the stocks with incomplete records during this period, which leaves 431 stocks remaining. For each stock, we download 8 news that are labelled to this stock every day through Yahoo Finance API\footnote{https://pypi.org/project/yfinance/}. 
Table \ref{tab:dataset} compares existing open-sourced financial news datasets with the SPNews dataset. 

The SPNews dataset serves as a valuable resource for the development of dynamic financial relationship graphs, presenting several noteworthy advantages. Firstly, it maintains a targeted focus on stocks by exclusively featuring news pertaining to companies within the SP500 Index, thus eliminating the presence of irrelevant information. Secondly, the dataset incorporates timestamps in the collection of news entries, facilitating a temporal analysis that is pivotal for capturing the temporal evolution of financial relationships. Lastly, the dataset annotates companies associated with each news item, providing a structured framework for the construction of inter-company relationships in financial modelling and analysis. 


\section{Experiment Setup}
In this section, we present experimental results that convey the effectiveness of the proposed FRI matrix. In order to verify the utility of the FRI framework, we conducted downstream experiments on the same set of graphs. The effectiveness of the FRI framework can be demonstrated by comparing the experimental results on downstream tasks and the FRI metric.

\subsection{Graph Dataset Construction}
In our experiment, we implemented five methods to construct the financial relationship graphs:
\begin{itemize}
\item $Static Graph$: Construct a relationship graph at the beginning of $T$ and the graph remains constant during the entire period,  i.e. $[\mathcal{G}_0=\mathcal{G}_1=...=\mathcal{G}_T]$. In our experiment, $T=236$ trading days. 
\item $Dynamic Graph Corr$: The edges between nodes is determined according to the value of each element of the correlation matrix \cite{xiang2022temporal}. 
\item $Dynamic Graph SPNews_{\tau=0}$ (Ours): The edges between nodes are built based on their co-occurrence in the \textit{SPNews} dataset. $\tau=0$ means we build an edge between two companies if they co-occurred in any news at least once on day $t$.
\item $Dynamic Graph SPNews_{\tau=1}$ (Ours): $\tau=1$ means we build an edge between two companies if they co-occurred in any news \emph{more than once} on day $t$.
\item $Dynamic Graph SPNews_{\tau=2}$ (Ours): $\tau=2$ means we build an edge between two companies if they co-occurred in any news \emph{more than twice} on day $t$.
\end{itemize}
\subsection{Downstream Task}
We select the stock trend prediction as the downstream task and regard it as a three-class classification task. The training set, validation set, and test set are distributed in a ratio of 8:1:1. 
For a node $\mathcal{V}(i)$ in graph $\mathcal{G}_t$, the label $y_i^t$ is 
\begin{align}
y_i^t= 
\begin{cases}
    negative,  & \text{if $r_i^{t+1} < -\mathrm{std}(r_i)$} \\
    neutral,  & \text{if $-\mathrm{std}(r_i) < r_i^{t+1} < \mathrm{std}(r_i)$} \\
    positive, & \text{if $r_i^{t+1} > \mathrm{std}(r_i)$}
\end{cases}
\end{align}%
where $r_i^{t+1}$ represents the rate of return of node $\mathcal{V}(i)$ on day $t+1$, $r_i$ represents the time series of node $\mathcal{V}(i)$'s returns during the whole dataset period, and $\mathrm{std}(r_i)$ represents the standard deviation of the time series $r_i$. The benefit of labelling three classes is that, from the investment management point of view, the investors focus more on the firms with large positive or negative returns because these firms have room for profit. Slightly positive or negative returns are usually regarded as normal fluctuations.

We select the Graph Attention Network (GAT) model as the baseline model, which is commonly used in graph-based stock trend prediction tasks \cite{cheng2021modeling,xiang2022temporal}. Inspired by \cite{li2021modeling}, we add a Long-Short Term Memory (LSTM) on top of GAT to encode the historical information as the node embedding. Following previous research \cite{cheng2021modeling}, we implement a 2-layer GAT followed by a multi-layer perceptron(MLP) as the classifier to get the classification results of the nodes. The $softmax$ activation function is used for the last layer. We use cross-entropy as the loss function which is formulated as follows:
\begin{align}
\mathcal{L} = - \sum_{i\in \mathcal{V}}\sum_{k=1}^K ~y_i^t(k)~\log(\hat{y}_i^t(k))
\end{align}%
where $K$ represents the number of classes, $k$ represents individual label $k$. $y_i^t(k)$ indicates whether the node $i$ belongs to label $k$ at time t. 
$\hat{y}_i^t(k)$ is the model predicted probability of node $i$ belonging to label $k$. The Adam \cite{kingma2014adam} optimizer is used to update model parameters. We keep the model architectures and hyper-parameters constant so that the model performance can represent the difference in the relationship graphs.

The accuracy (ACC) and Macro F1 score are adopted as evaluation metrics of models. Macro F1 score is defined as the mean of F1 scores of each class:
\begin{align}
Macro F1 = \frac{1}{K} \sum_{k=1}^K F1\_score
\end{align}%
where $K$ represents the number of classes, $k$ represents individual label $k$.

\section{Results and Discussion}
We evaluate the constructed relationship graphs using both the \textit{FRI framework} and the downstream task performance.
\subsection{Result Comparison}

\begin{table}[]
\centering
\begin{tabular}{llllll}
\hline
                                   & CSS    & AECR   & $\Delta \beta$  & $\Delta DCC$ \\ \hline
Static Graph                       & 0.383 & 0.078          & 0            & 0            \\
$Dynamic Graph Corr$         & 0.411 & 0.359 & 0.026          & -0.031       \\
$Dynamic Graph SPNews_{\tau=0}$ & \textbf{0.476} & \textbf{0.616} & \textbf{0.071}        &\textbf{0.429}\\
$Dynamic Graph SPNews_{\tau=1}$ & 0.448 & 0.563 & 0.049           & 0.148     \\ 
$Dynamic Graph SPNews_{\tau=2}$ & 0.429 & 0.522 & 0.060           & -0.350     \\ \hline
\end{tabular}
\caption{Financial relationship graph comparison in FRI matrix. \textbf{Bold} shows the best results.}
\label{tab:graph-fri-comparison}
\end{table}

\textbf{Table \ref{tab:graph-fri-comparison}} presents the evaluation results of different graphs under the FRI framework. Notably, the static graph exhibits poor overall performance, with its various assessment metrics notably inferior to those of dynamic graphs. Specifically, the \textit{CSS} of the static graph (0.383) demonstrates the smallest gap from other dynamic graphs (with a minimum value of 0.411). This is mainly because \textit{CSS} is the average value of $CS_{\mathcal{G}_t}$ of each graph $\mathcal{G}_t$, and for static graphs, $\mathcal{G}_0 = \mathcal{G}_1 = ... = \mathcal{G}_T$. Therefore, the small difference between the \textit{CSS} of the static and dynamic graphs proves that $\mathcal{G}_0$ does effectively capture the real relationships that exist between certain companies.

Aside from the \textit{CSS} indicator, the other three indicators vertically evaluate the graph set $\mathcal{G}$, that is, we evaluate whether the edges between each pair of companies effectively reflect the changes in the relationship over time. Therefore, the performance of static graphs on the remaining three indicators is much lower than that of dynamic graphs. Among them, the \textit{AECR} indicator is only $0.0785$, which has the largest difference with dynamic graphs. This means that among all the edges in the static graph, only $7.85\%$ of the edges correspond to events that cause large correlation fluctuations. However, among the four dynamic graphs, the lowest value is $35.97\%$ and the best performing graph has an AECR of $61.63\%$. At the same time, we can find that the \textit{AECR} indicators of the SPNews-based dynamic graphs are significantly higher than those of the correlation-based dynamic graph. 
Since all connected node pairs have the same density of edges in the static graph set, $\Delta \beta$ and $\Delta DCC$ indicators are not applicable to it. 
It is worth noting that $Dynamic Graph Corr$ has poor explanatory power for return correlations, suggesting that changes in correlations are necessary but not sufficient for the existence of true relationships. 

\textbf{Table \ref{tab:graph-model-comparison}} presents the results of conducting the downstream task on each graph. By comparing Table \ref{tab:graph-fri-comparison} and Table \ref{tab:graph-model-comparison}, we can find that the conclusions drawn by the two evaluation methods are not exactly the same. However, if we sort all graphs according to the statistics in the tables, we can get the following results:
\begin{itemize}
\item Table \ref{tab:graph-fri-comparison}: ${\tau=0} > {\tau=1} > {\tau=2} > Corr > Static Graph$ 
\item Table \ref{tab:graph-model-comparison}: ${\tau=1} > {\tau=2} > {\tau=0} > Corr > Static Graph$ 
\end{itemize}
From the sorting results, we can see that the results of FRI framework and downstream task evaluation are in the same trend. 
We don't suggest that our proposed FRI framework should replace downstream task evaluation, but rather that it can augment the decision making process, which is particularly useful if downstream evaluation is computationally expensive.
The framework serves as a guide to which graphs are likely to perform well in downstream tasks. We believe that the interpretability and evaluation of graphs is a direction worth exploring, and this work may inspire future researchers to think beyond downstream tasks when evaluating financial graphs.


\begin{table}[]
\begin{tabular}{lcc}
\hline
                                   & ACC   & Macro F1 \\ \hline
$Static Graph$                       & 0.345 & 0.241    \\
$Dynamic Graph Corr$          & 0.347 & 0.244    \\
$Dynamic Graph SPNews_{\tau=0}$ & 0.393 & 0.264    \\
$Dynamic Graph SPNews_{\tau=1}$ & \textbf{0.459} & \textbf{0.298}    \\ 
$Dynamic Graph SPNews_{\tau=2}$ & 0.401 & 0.267    \\ \hline
\end{tabular}
\caption{GAT model performance comparison using different graphs. \textbf{Bold} shows the best results.}
\label{tab:graph-model-comparison}
\end{table}


\subsection{Discussion}
In this section, we provide a discussion aimed at enhancing the readers understanding of the FRI framework. To do this, we present a case study and an examination of the framework's limitations.

\begin{figure*}[]  
    \centering
    \includegraphics[scale = 0.45]{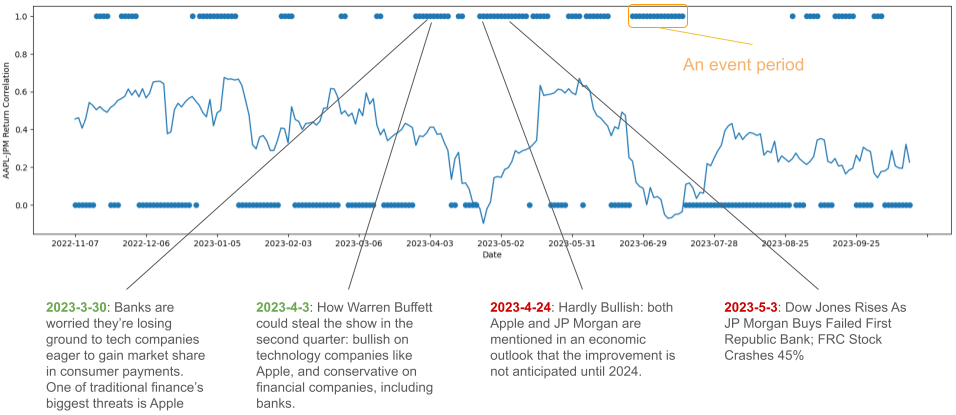}
    \caption{Apple-JP Morgan Case Study. Line plot: the rolling return correlation. Scatter points: edge index $\mu_t^{(AAPL, JPM)}$.}
    \label{fig:case-study}
\end{figure*}

\textbf{Case Study} To provide readers with a deep understanding of the FRI framework, we conducted a case study focusing on the Average Event Capturing Rate (ACER). Figure \ref{fig:case-study} illustrates the variations in the rolling 21-day return correlation between Apple Inc and JPMorgan Chase \& Co over the entire period covered by our dataset. The scatter points distributed along the vertical axis at 0 and 1 represent whether an edge exists between Apple and JP Morgan in our relationship graph at time $t$. For instance, if an edge exists between these two companies in our graph $\mathcal{G}_t$ on day $t$, i.e. $\mu_t^{(AAPL, JMP)} = 1$, then a point will be plotted at the position $(t, 1)$ in Figure \ref{fig:case-study}. Conversely, if $\mu_t^{(AAPL, JPM)} = 0$, a point will be plotted at the position $(t, 0)$ in the figure. A continuous series of scatter points at the vertical coordinate of 1 represents an \textit{event period}. We observe that periods in which the return correlation fluctuates significantly generally occur within our event period, aligning with our intuition and assumptions. In contrast, correlations tend to experience smaller fluctuations outside of event periods. For example, between March 2023 and April 2023, there was a sharp decline in the return correlation between Apple and JP Morgan. During this period, our SPNews dataset contains news items that may have led to divergent stock trends for these two companies, thereby weakening their correlation. Consequently, the Event Capture (EC) within this event period is assigned a value of 1. In summary, the Average Event Capturing Rate (AECR) serves as a metric to assess significant changes in return correlation captured by edges in the relationship graph.

\textbf{Limitation} Our method has a few limitations which we plan to explore further in future work. First, the \textit{FRI} framework mainly focuses on dynamic graph evaluation, in which three indicators are not applicable to the static graphs. Second, the FRI method cannot evaluate the quality of the weight of each edge ($\nu_t^{(A,B)}$ in this paper) in the weighted graph. Therefore, the FRI method cannot be used to compare methods for calculating edge weights.

\section{Conclusion}
In conclusion, this paper released a financial news dataset SPNews which enables the construction of various financial entity relationship graphs. Besides, we proposed a novel financial relationship graph evaluation framework that is independent of downstream tasks and models. The FRI framework addresses a gap in the literature where relationship graph evaluations often rely heavily on the model and downstream task. Experimental results demonstrate the utility of the FRI framework and also prove that graphs constructed based on SPNews are better than graphs constructed based on returns correlation, thereby proving the value of the SPNews dataset.



\bibliographystyle{ACM-Reference-Format}
\bibliography{sample-base}

\end{document}